\documentclass[wsdraft]{ws-procs11x85}

\newif\ifpreprint
\preprinttrue

\usepackage{ws-procs-thm}           

\usepackage{savesym}
\usepackage{amsmath}
\savesymbol{captionbox}
\usepackage{subcaption}
\usepackage{graphicx}

\begin{document}
\pagenumbering{arabic}
\title{Implicitly and Differentiably Representing Protein Surfaces and Interfaces}

\author{Cory B. Scott$^\dag$, Charlie Rothschild, and Benjamin Nye}

\address{Department of Mathematics and Computer Science, Colorado College,\\
Colorado Springs, CO 80909, USA\\
$^\dag$E-mail: \email{cbs@coloradocollege.edu}
\url{https://sites.coloradocollege.edu/cbs/}}

\thispagestyle{empty}

\begin{abstract}
We introduce a pipeline for representing a protein, or protein complex, as the union of signed distance functions (SDFs) by representing each atom as a sphere with the appropriate radius. While this idea has been used previously as a way to render images of proteins, it has not, to our knowledge, been widely adopted in a machine learning setting. Mirroring recent successful work applying SDFs to represent 3D geometry, we present a proof of concept that this representation of proteins could be useful in several biologically relevant applications. We also propose further experiments that are necessary to validate the proposed approach.
\end{abstract}


\ifpreprint
\copyrightinfo{\copyright\ 2025 The Authors. Preprint of an article submitted for consideration in Pacific Symposium on Biocomputing © 2026 [copyright World Scientific Publishing Company] [https://psb.stanford.edu/psb-online/] }
\fi


\section{Introduction}
Modeling the three-dimensional shape of proteins is crucially important for understanding how they interact with other molecules.
Advances in representations of protein structure have lead to corresponding advances in the ability for machine learning (ML) methods to predict biologically relevant properties of proteins, such as: binding affinity with bioactive molecules; prediction of protein conformation; and simulation of protein behavior under a variety of external conditions. One common representation of proteins that has received much attention is the \emph{protein surface} \cite{connoly1983sas}, also sometimes called the \emph{solvent accessible surface} (SAS), which is a 2D manifold that represents the portion of the protein's surface that is physically accessible to a solvent.

In this work, we propose an alternate protein surface representation: an isosurface of a signed distance function (SDF), produced by smoothing the boolean union of individual spherical SDFs for each of the protein's component atoms. This approach naturally produces the same notion of a protein's solvent-accessible surface. We present a proof-of-concept demonstration of how this representation can be applied to predict protein-protein interactions; further work is needed to validate the proposed approach. The main contributions of this work are as follows: 1) we discuss prior work training machine learning models on protein surface geometry; 2) we demonstrate a way to represent this surface as the zero-level set of a signed distance function, constructed with boolean operations; 3) we investigate using acceleration structures to query a protein SDF more efficiently; 4) we produce a dataset of protein-protein interface meshes, and provide code to reproduce our results.

\subsection{Mathematical and Biological Background}
We first introduce some necessary background in protein biology, geometry, and geometric machine learning. Throughout this paper, variables with an arrow ($\Vec{x}$) represent points in $n$D Euclidean space, lowercase letters represent constants, and $|| \cdot ||$ is the $n$D Euclidean norm. 
\paragraph{Signed Distance Functions (SDFs)}
Let $\Omega \subset \mathbb{R}^n$ be a compact subset of Euclidean space, and let $\partial\Omega$ represent its boundary. For any point $\Vec{x} \in \mathbb{R}^n$, the signed distance $d_\Omega(\Vec{x})$ is  defined as: 

\begin{equation}
    \label{eqn:sdf_defn}
    d_\Omega(\Vec{x}) = \begin{cases} 
      0 & x \in \partial\Omega \\
      -|| \Vec{x} - \Vec{y} || & \Vec{x} \in \Omega \\
      || \Vec{x} - \Vec{y} || & \Vec{x} \notin \Omega \\
   \end{cases} \\
   \text{where} \quad  \Vec{y} = \arg \min_{\Vec{y} \in \partial\Omega} || \Vec{x} - \Vec{y} || 
\end{equation}

In other words, the signed distance measures the distance between any point and the boundary of $\Omega$, with the sign indicating whether a point is inside the shape or not (some authors take the opposite sign as convention, i.e. positive values denote an object's interior). Signed distance functions have many properties that make them useful in a machine learning context: 1) they have unit gradient everywhere the gradient is defined; 2) analytic formulae have been found for a wide variety of SDFs for specfic shapes \cite{quilez2008modeling}; and most significantly for this work, 3) simple SDFs can be combined into SDFs for more complex shapes using basic boolean operations. We will specifically make use of the exponential \emph{smooth-min} operation, one of a family of SDF blending operations originally proposed by Quilez \cite{quilez2013smooth}. If $d_1(x), d_2(x), \ldots d_n(x)$ are a set of SDFs, then their smooth-min $\mathbf{d}$ is given by 
\begin{equation}
    \label{eqn:smooth_min}
    \mathbf{d}(\Vec{x}) = -k \log \left( \sum_{i=1}^n e^{-\frac{1}{k} d_{i}(\Vec{x})} \right) \\
    \text{where $k$ determines the smoothing radius.}    
\end{equation}

As a relevant example of a specific closed-form analytic SDF, the SDF $d$ for a $n$-dimensional sphere of radius $r$ centered at $\Vec{y}$ is given by $d(\Vec{x}) = ||\Vec{x} - \Vec{y}|| - r $. 

When SDFs are combined with boolean operations, the resulting function may not be a true SDF, in the sense that it may not satisfy Equation \ref{eqn:sdf_defn}. In practice boolean operations produce distance fields that are approximate enough for applications like raycasting or training ML models.

\paragraph{Protein Surfaces}
Protein surfaces \cite{shrake1973environment} are a common way to represent the parts of a protein that are available to react/bind with other proteins or small molecules. A protein surface is typically calculated by representing each atom as a hard-shell sphere of a given radius, and rolling a simulated spherical probe (typically taken to have the radius of a single hydrogen atom, 1\r{A}) over the collection of atoms.
The final mesh representing the protein is produced by discretizing the surface traced by the probe. If we trace the center of the probe sphere, we get the Solvent Accessible Surface (SAS) of the protein. Tracing instead all the contact points between the probe and the atoms of our molecule produces the Solvent Excluded Surface. See Figure \ref{fig:multiple_proteins} for examples of protein solvent excluded surfaces. 
\subsection{ML Analysis of Protein Surfaces}
In this section we briefly discuss prior work that applies machine learning to the task of protein surface analysis. Protein design and analysis is a rapidly evolving application area of machine learning \cite{casadio2022machine, notin2024machine}. For a more thorough review of ML for protein design and concepts in geometric machine learning, we refer the reader to Cheng et al. \cite{cheng2008machine} and Bronstein et. al \cite{bronstein2017geometric} respectively. Protein surfaces have been widely adopted as a representation of proteins that facilitates training machine learning models to identify possible protein-protein and protein-ligand interactions \cite{ bordner2007protein, gainza2020deciphering, atz2021geometric, mendez2021geometric, mylonas2021deepsurf}. While much of the prior work on protein surfaces has focused on meshes as an intermediate representation, there is a complimentary vein of work that uses signed distance functions to represent proteins. SDFs have been thoroughly examined in the context of rendering 2D and 3D images of proteins \cite{parulek2012implicit,parulek2013fast,klepavcrendering}. Much of this prior work uses machine learning or an SDF representation of proteins, but not both. A notable exception is Sverrisson et al. \cite{sverrisson2021fast} which uses a similar atomic union SDF to the one we propose. However, that work mainly examines the definition of a convolution-like operation on implicitly defined protein surfaces, and does not consider combinations of such surfaces with constructive solid geometry operations as we do in the present work.

\section{Method}
\subsection{Protein Representation}
We propose representing a protein as the union of spherical SDFs of each of its component atoms, where each atom is a sphere sized according to its van der Waals radius (as reported in \cite{bondi1964van}). These spherical SDFs are combined using the smooth-min operation, which we and others have observed in practice \cite{patane2015state,sverrisson2021fast} to resemble SAS and SES computed via other means. See Figure \ref{fig:multiple_proteins} for several examples of protein SDFs computed according to this method. We implement these SDFs in Pytorch \cite{paszke2017automatic}, enabling backpropagation of error through loss functions composed of protein SDF queries. 

\newcommand{\protsubfigwidth}{.23\linewidth}

\begin{figure}
    \centering

    \null \hfill
    \begin{subfigure}[b]{\protsubfigwidth}
        \includegraphics[width=\textwidth]{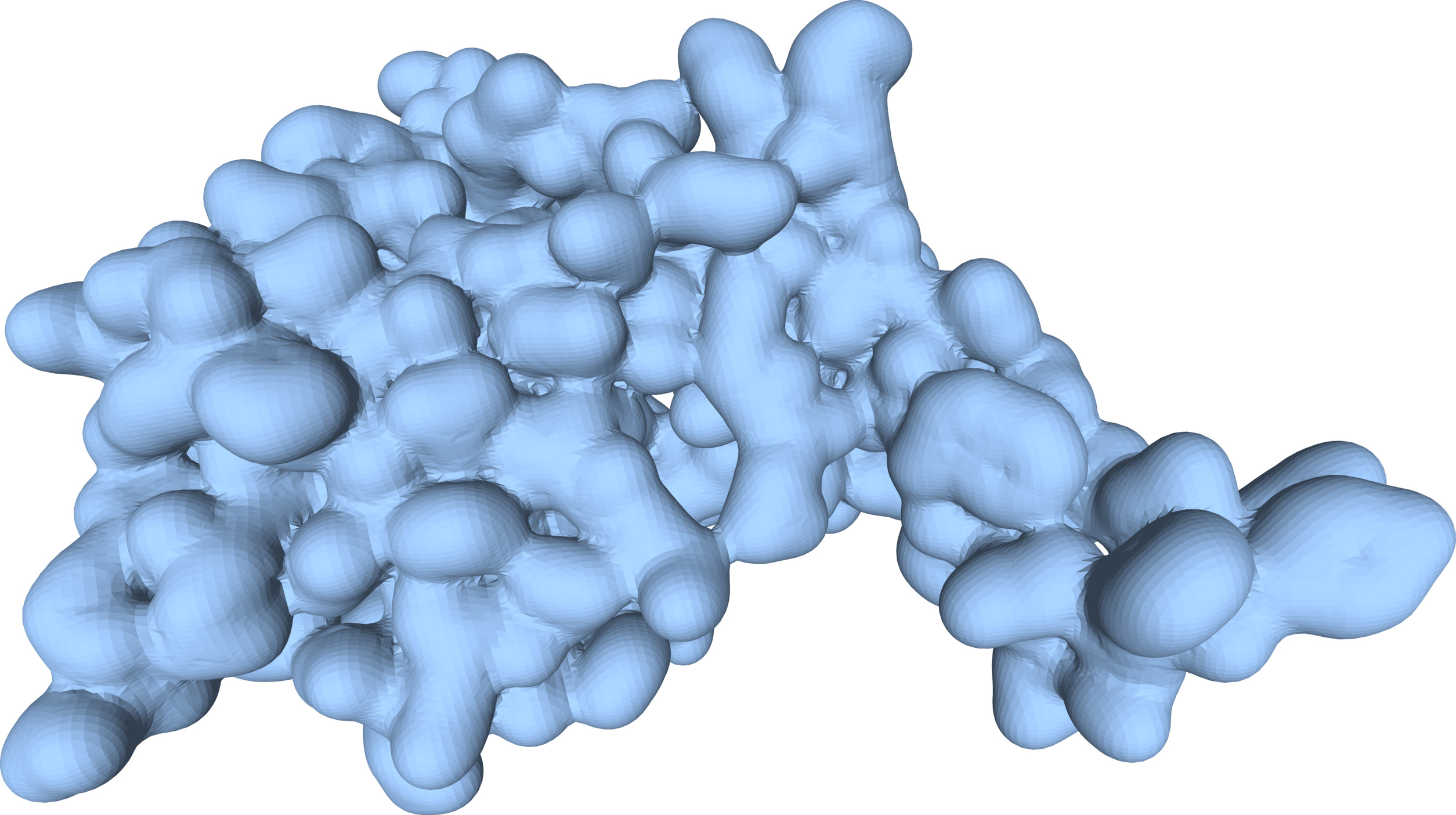}
        \caption{2I9B}
        \label{fig:multi_prot_2i9b}
    \end{subfigure}
    \hfil
    \begin{subfigure}[b]{\protsubfigwidth}
        \includegraphics[width=\textwidth]{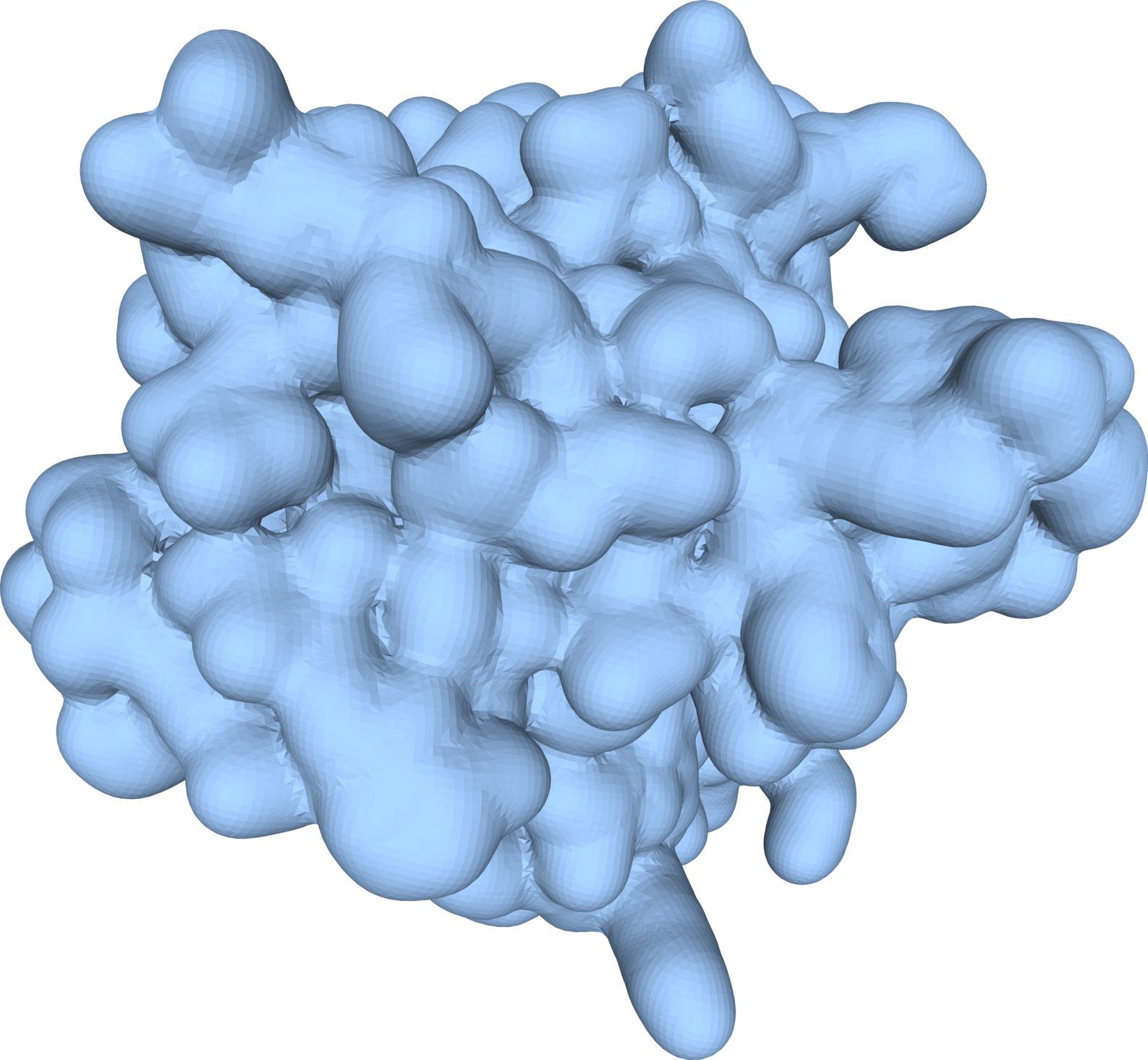}
        \caption{1JK9}
        \label{fig:multi_prot_1jk9}
    \end{subfigure} 
    \hfil
    \begin{subfigure}[b]{\protsubfigwidth}
            \includegraphics[width=\textwidth]{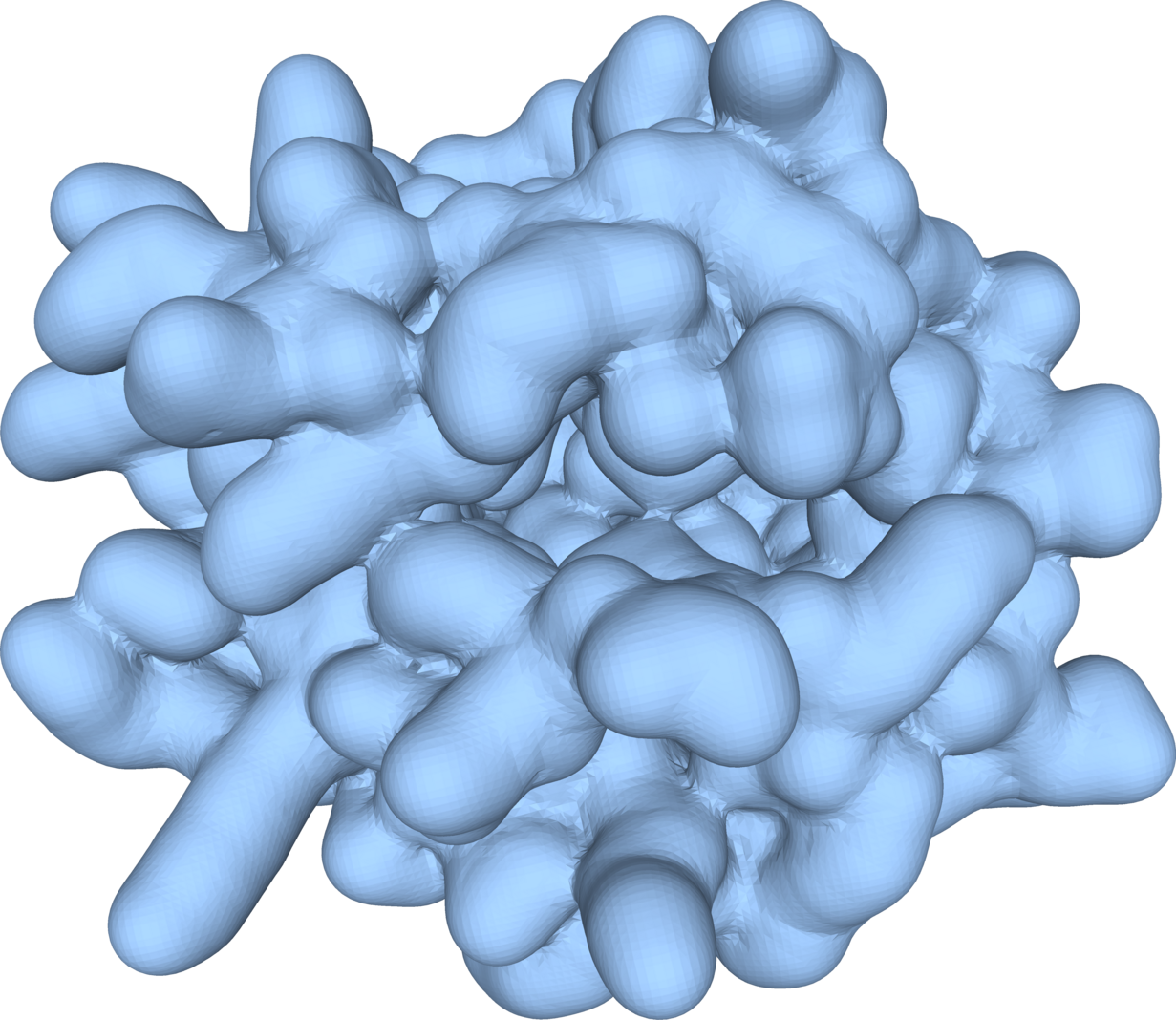}
            \caption{1WEJ}
            \label{fig:multi_prot_1wej}
    \end{subfigure}
    \hfil
    \begin{subfigure}[b]{\protsubfigwidth}
        \includegraphics[width=\textwidth]{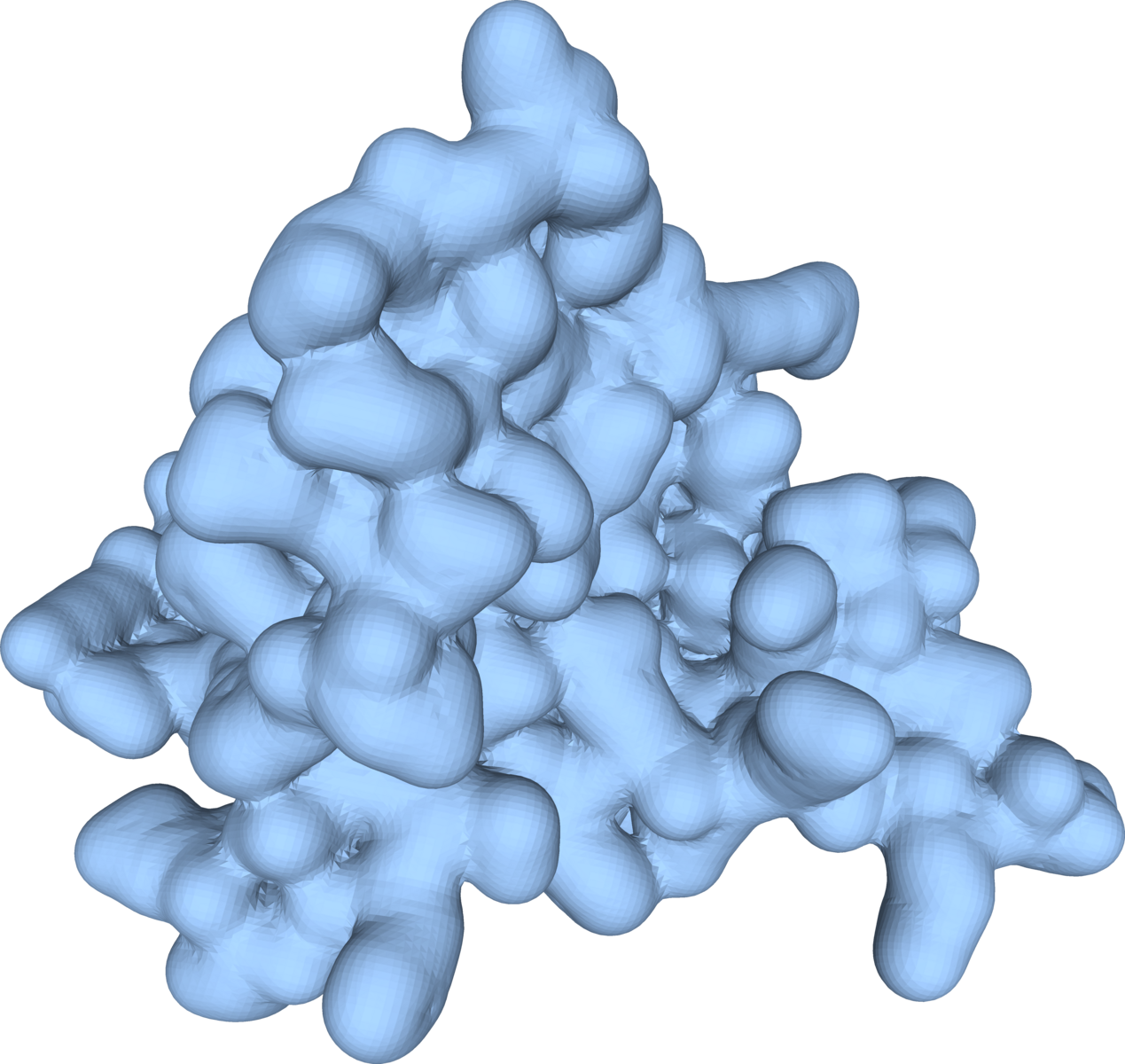}
        \caption{1PXV}
        \label{fig:multi_prot_1pxv}
    \end{subfigure} \hfill \null \\

    \null \hfill
    \begin{subfigure}[b]{\protsubfigwidth}
        \includegraphics[width=.65\textwidth]{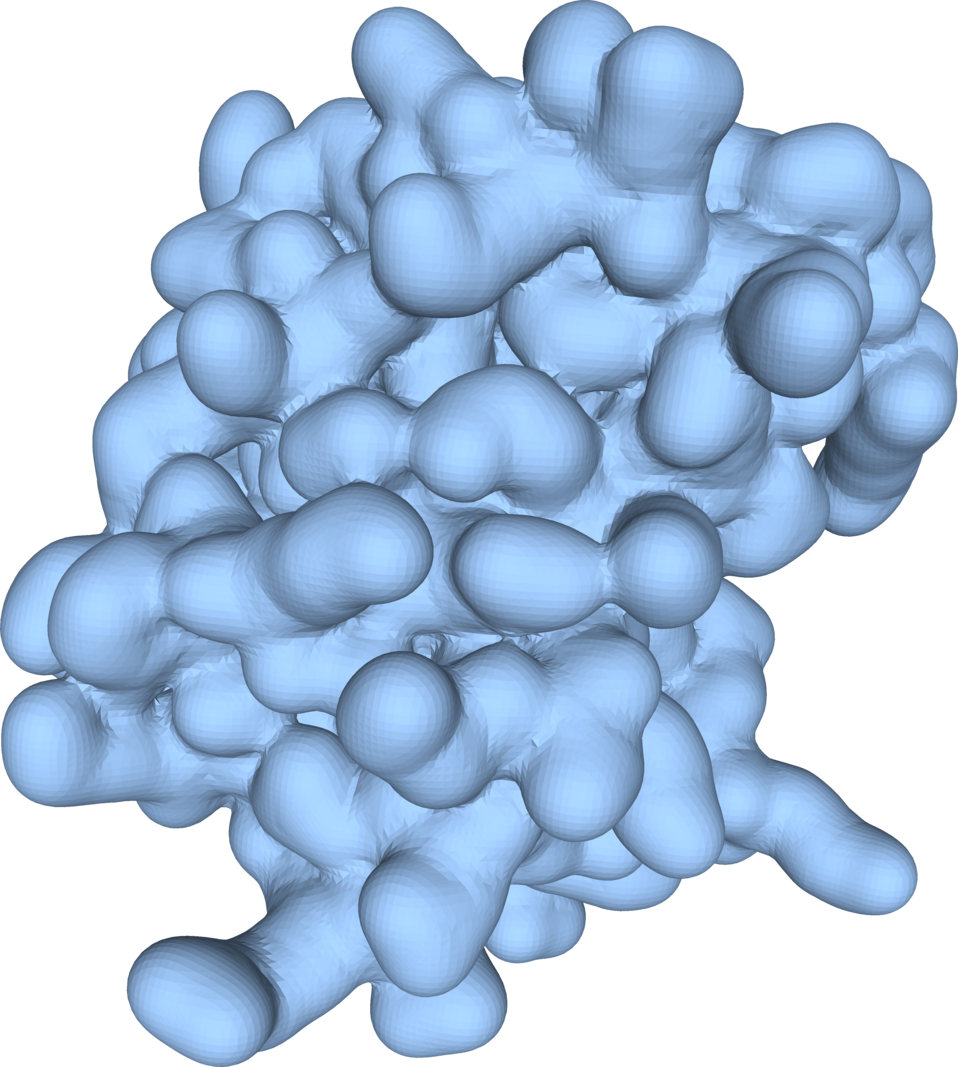}
        \caption{7CEI}
        \label{fig:multi_prot_7cei}
    \end{subfigure}
    \hfil
    \begin{subfigure}[b]{\protsubfigwidth}
        \includegraphics[width=\textwidth]{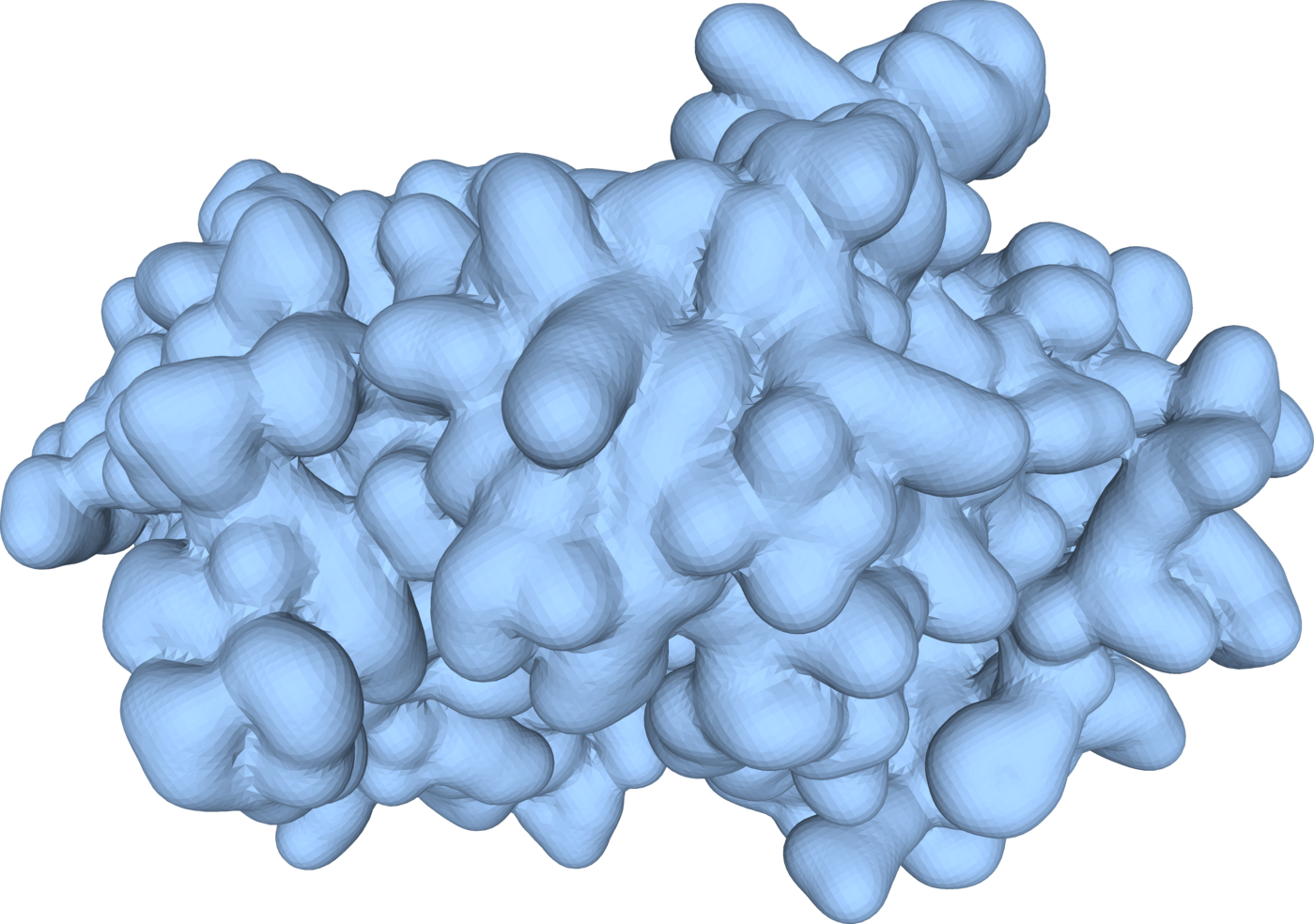}
        \caption{2GAF}
        \label{fig:multi_prot_2gaf}
    \end{subfigure} 
    \hfil
    \begin{subfigure}[b]{\protsubfigwidth}
        \includegraphics[width=\textwidth]{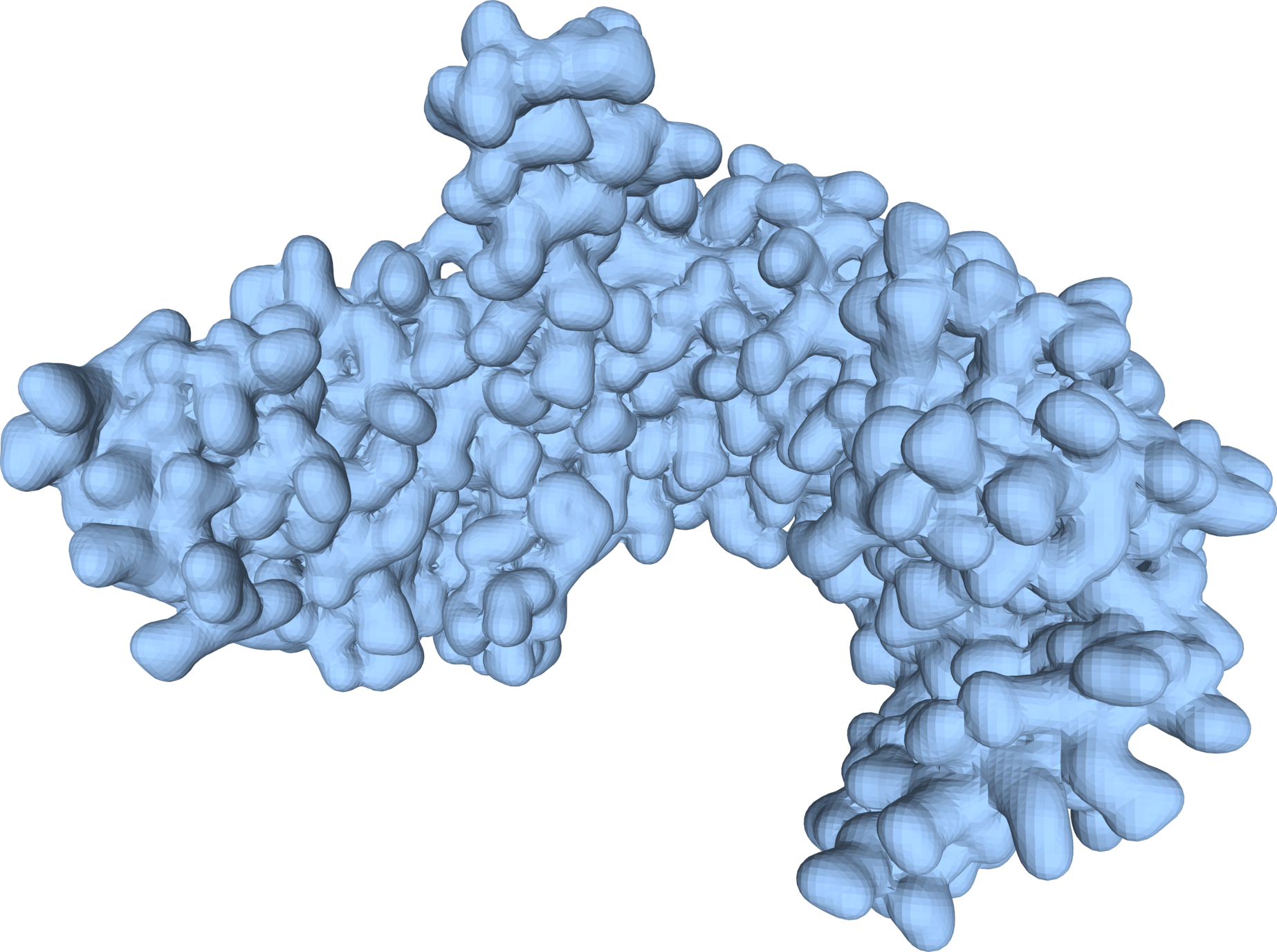}
        \caption{1IBR}
        \label{fig:multi_prot_1ibr}
    \end{subfigure}
    \hfil
    \begin{subfigure}[b]{\protsubfigwidth}
        \includegraphics[width=\textwidth]{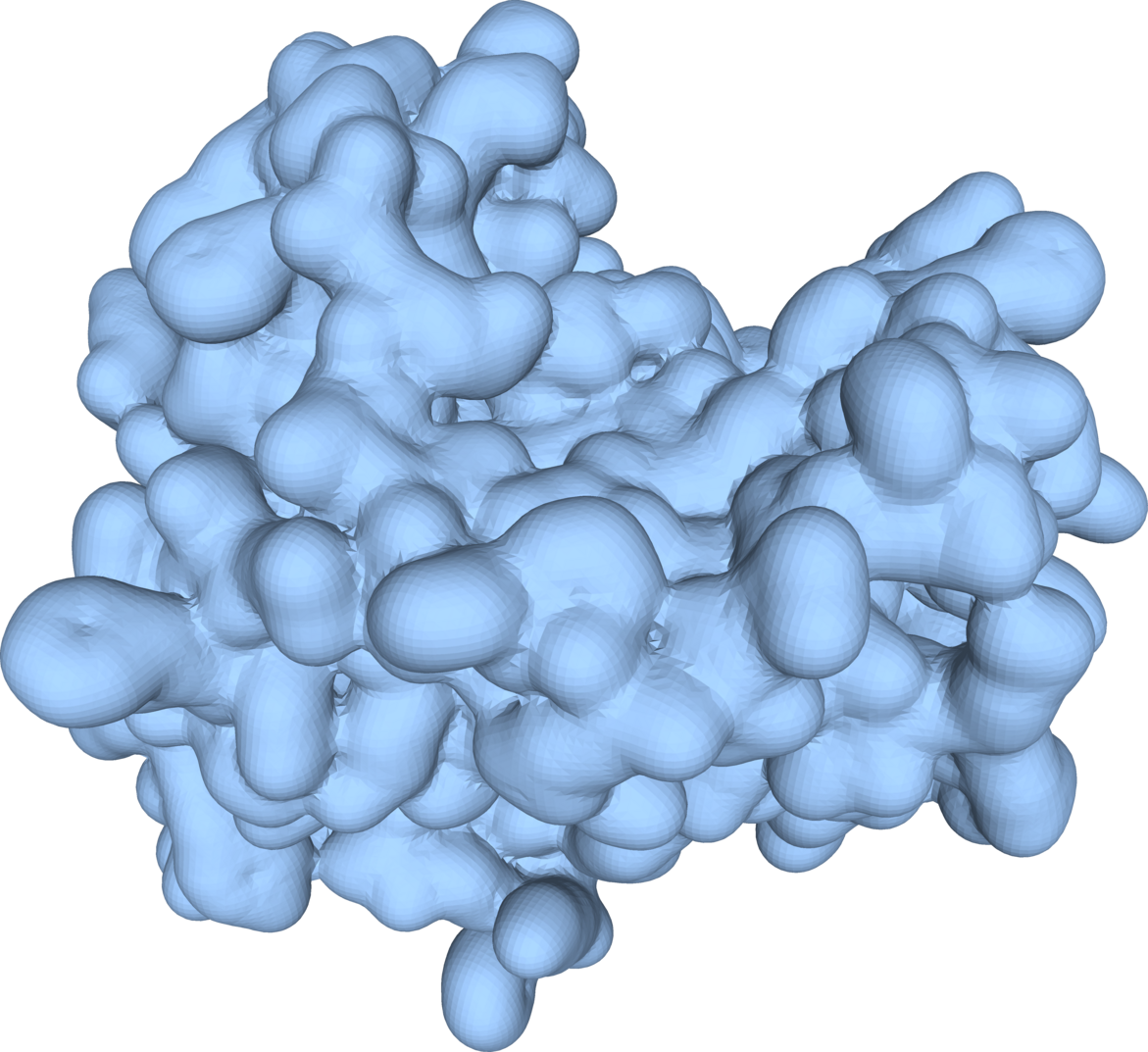}
        \caption{1A2K}
        \label{fig:multi_prot_1a2k}
    \end{subfigure}
    \hfill \null
    
    \caption{$d = 0$-level set meshes of multiple proteins from the DB5 dataset. Each protein imaged here is the ``left" chain of the protein complex, in its bound confirmation. }
    \label{fig:multiple_proteins}
\end{figure}

\begin{figure}
    \centering
    \includegraphics[width=.97\linewidth]{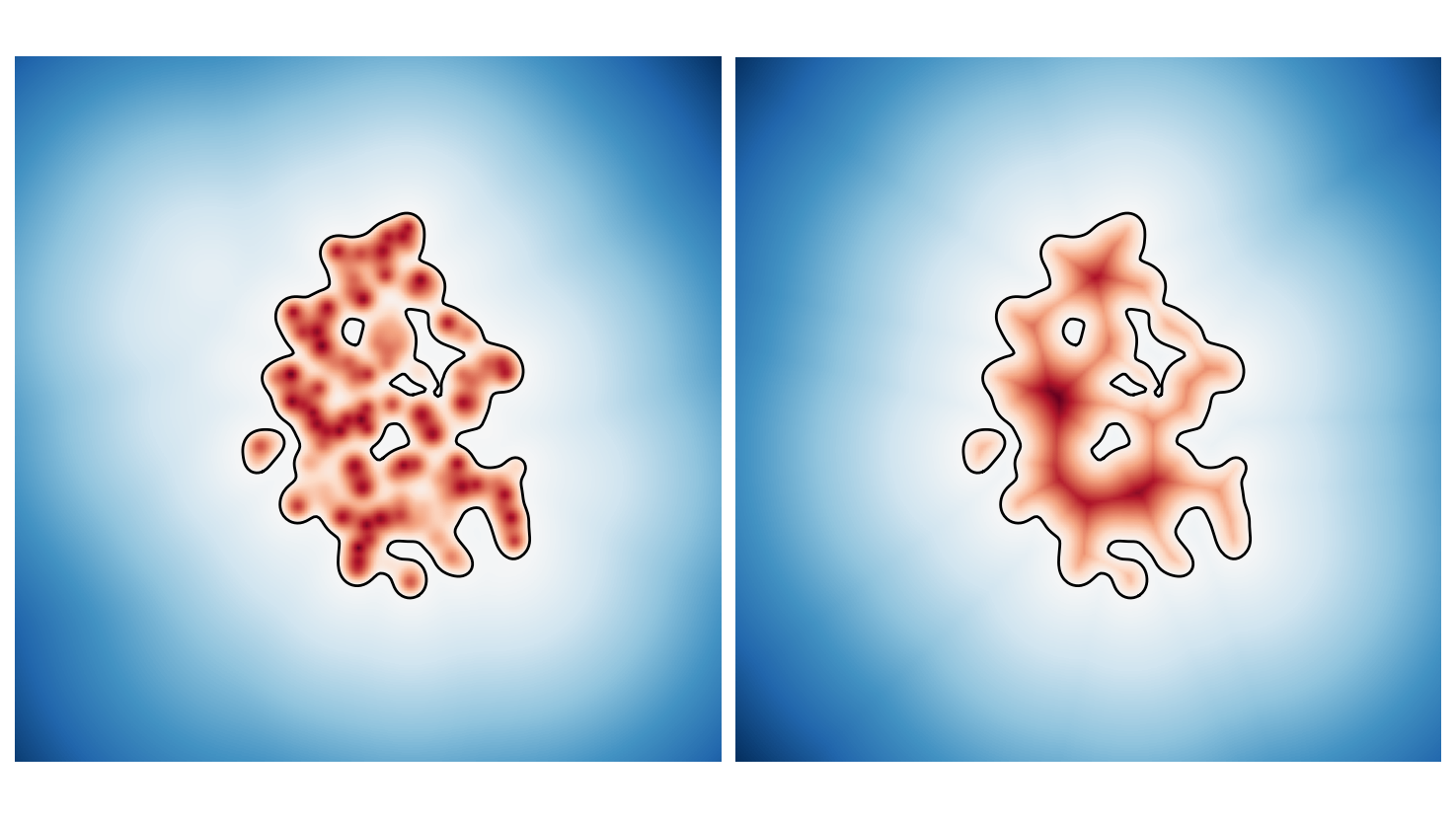}
    \caption{An illustration of one possible issue when representing a protein as a collection of spherical SDFs. The left shows the union SDF, while the right shows the true distance to the $d=0$ isosurface. The union SDF is incorrect in the protein's interior.}
    \label{fig:flat_prot_figures}
\end{figure}

\subsubsection{Possible Issues}
As noted above, boolean operations on SDFs are not guaranteed to produce a valid SDF. The effect of this on our proposed protein SDF can be seen in Figure \ref{fig:flat_prot_figures}. While the resulting SDF has the correct $d = 0$ isosurface which would be expected from performing the boolean operation, this technically results in a ``broken'' or incorrect SDF in the protein interior. The exterior distance values are still correct, but the interior values are instead a lower bound.

\subsection{Intersection of Protein Chains}
\begin{figure}
    \centering
    \begin{subfigure}[c]{.33\linewidth}
        \includegraphics[width=\linewidth]{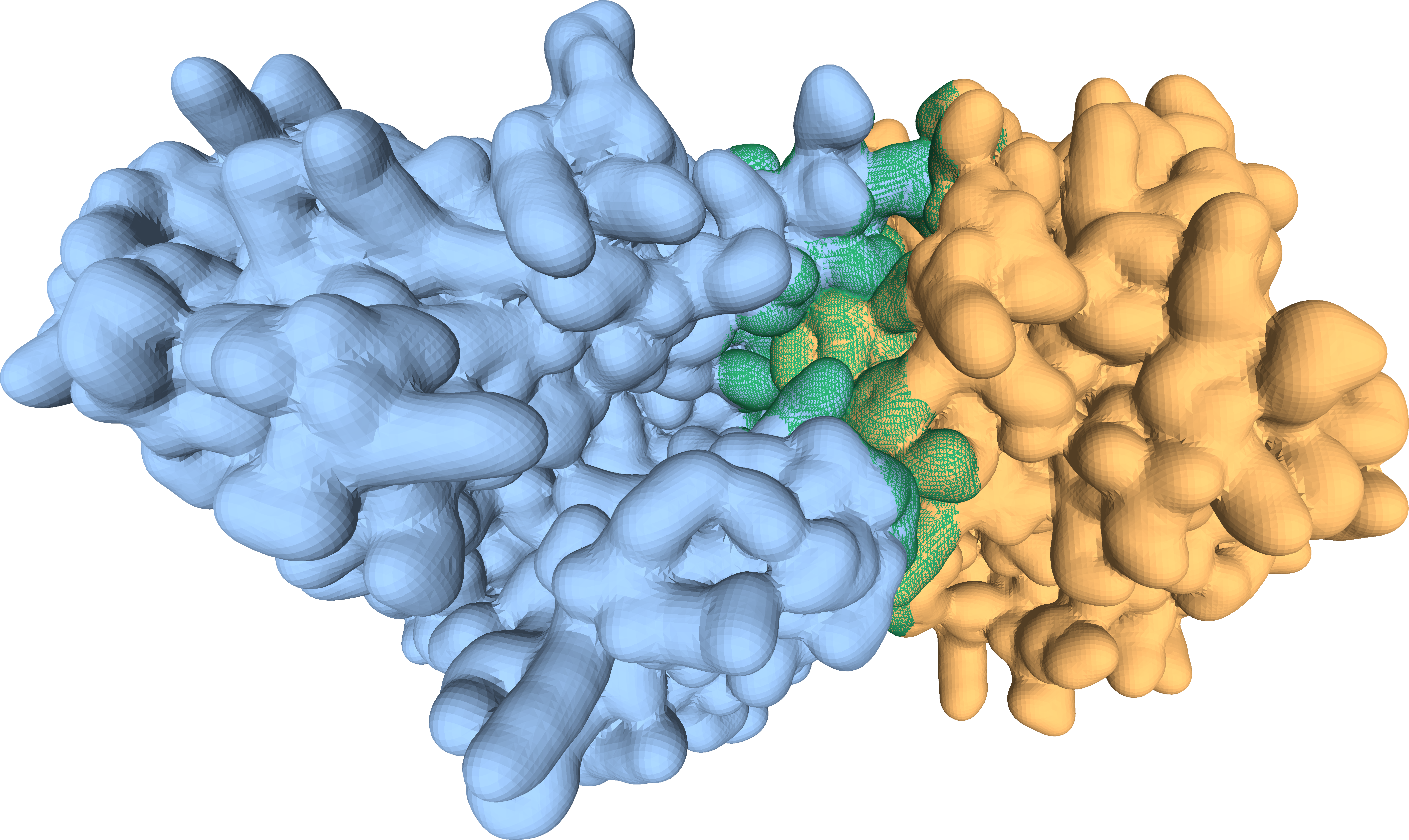}
        \caption{}
        \label{subfig:1wdw_both}
    \end{subfigure} 
    \hfill \begin{subfigure}[c]{0.16\linewidth}
        \includegraphics[width=\linewidth]{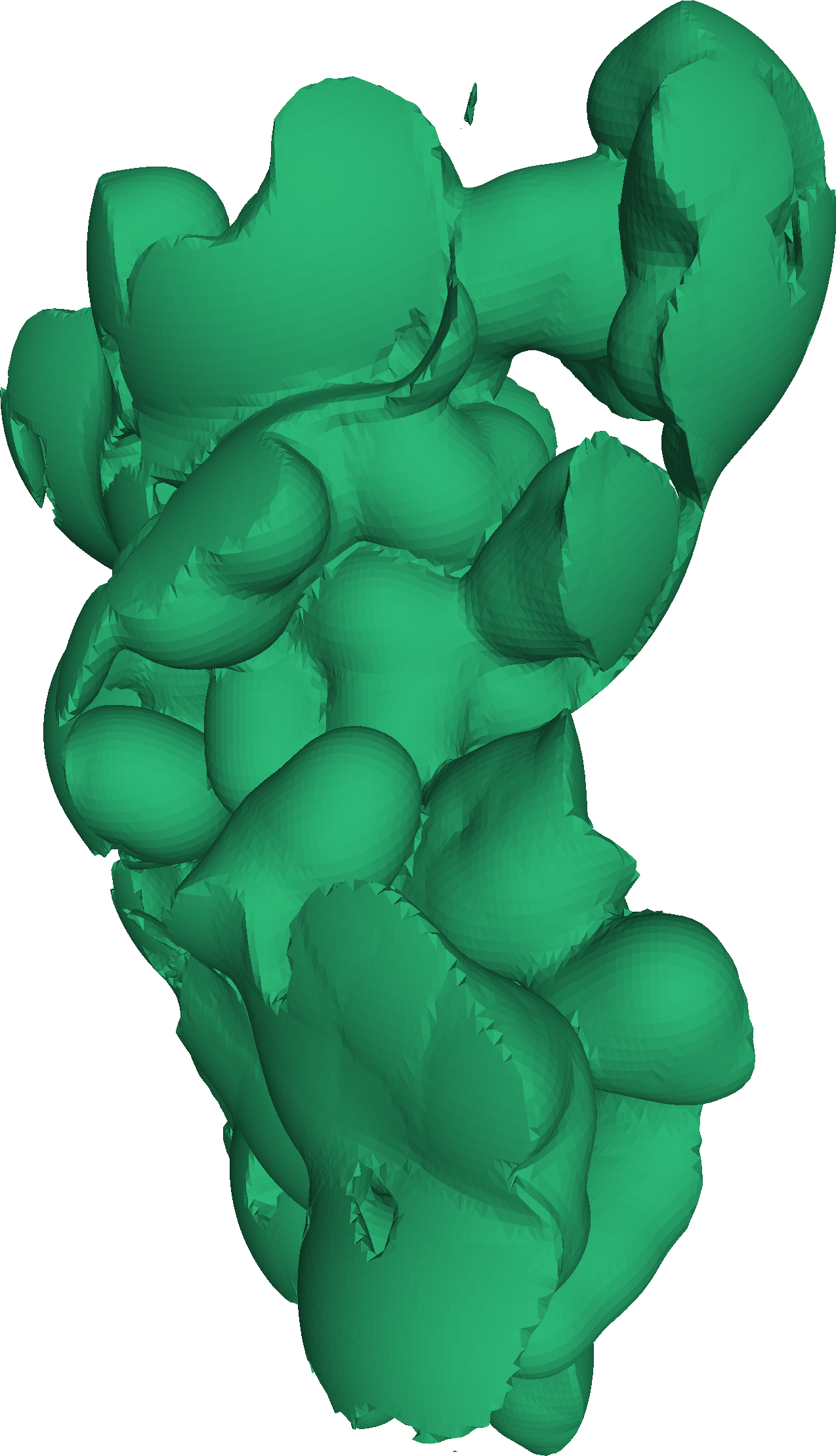}
        \caption{}
        \label{subfig:1wdw_inter_01}
    \end{subfigure}
    \hfill
    \begin{subfigure}[c]{0.33\linewidth}
        \includegraphics[width=\linewidth]{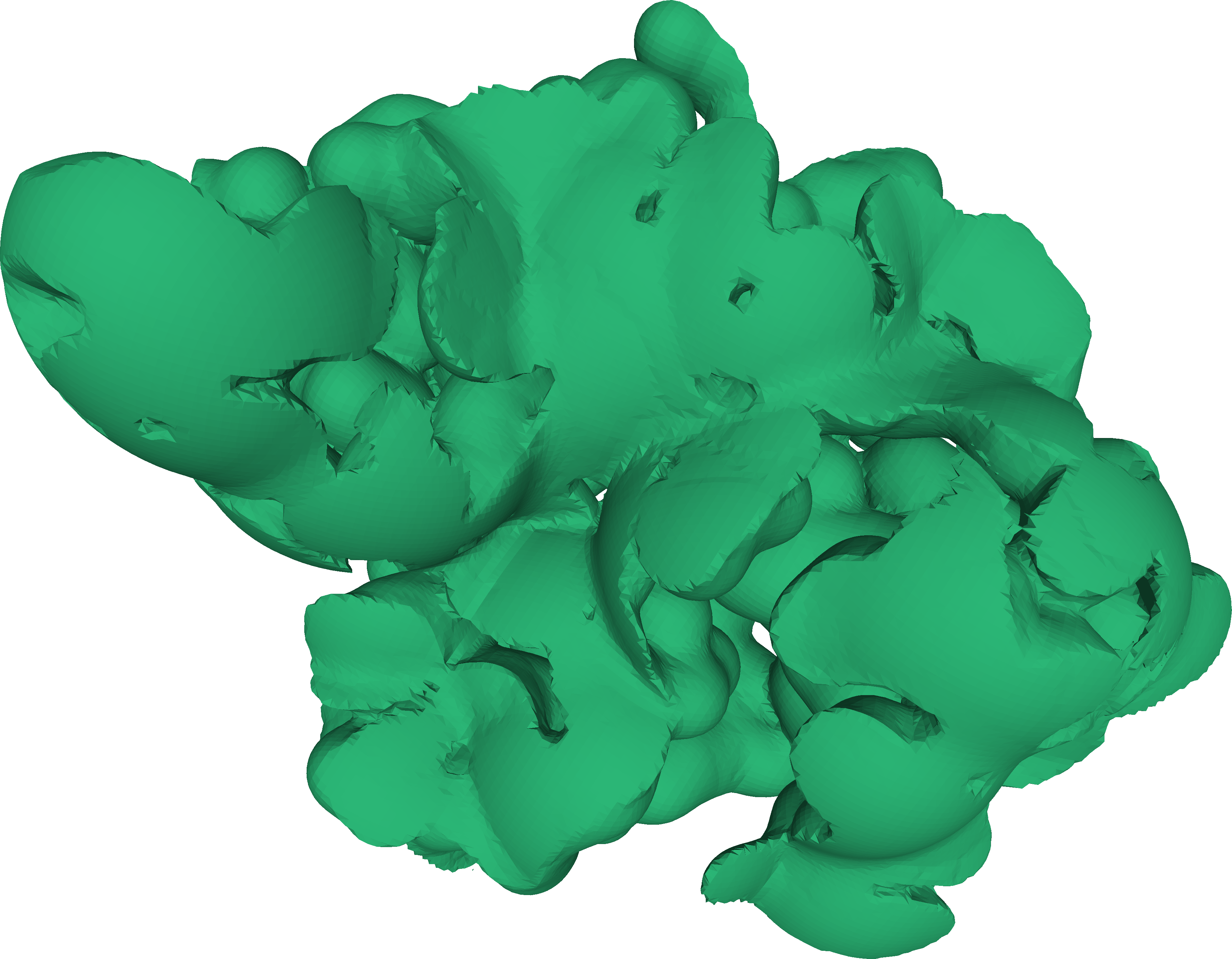}
        \caption{}
        \label{subfig:1wdw_inter_02}
    \end{subfigure}
    \hfill \null
    \caption{An example of the $d(\mathbf{x}) = 0$ isosurface generated by our method for the protein 1WDW from the DB5 dataset. Left: The intersection isosurface rendered alongside both chains of the protein complex (color denotes chain ID). Nodes in the intersection region have been highlighted in green. Right: two views of the isolated intersection mesh. }
    \label{fig:prot_intersect_ex1}
\end{figure}

\begin{figure}
    \centering
    \includegraphics[width=.5\linewidth]{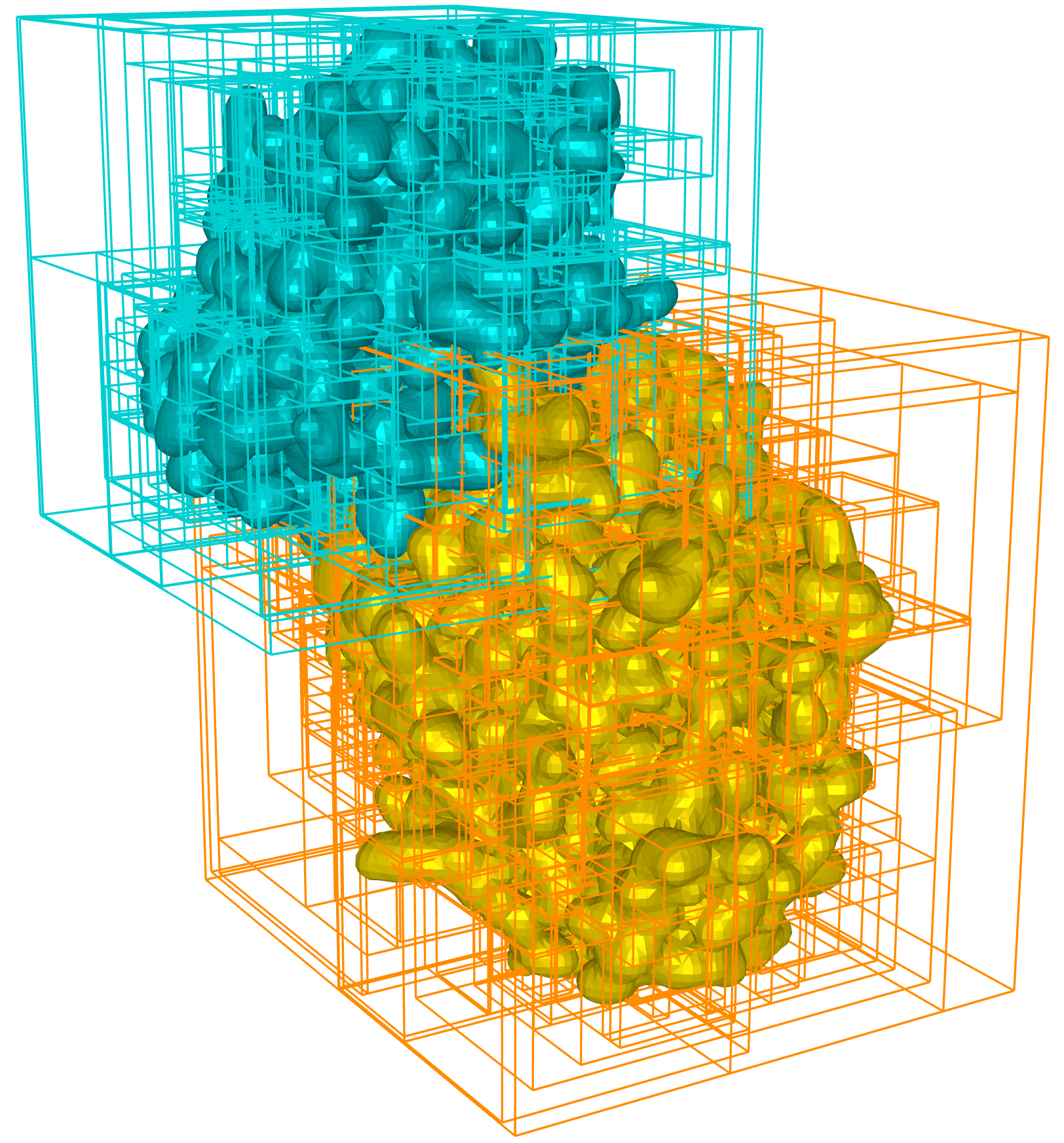}
    \caption{A visualization of the bounding volume hierarchy built by our method for the protein 1WDW.}
    \label{fig:bvh_with_bounding_boxes} 
\end{figure}

\begin{figure}
    \centering
    \null \hfill
    \begin{subfigure}[c]{.31\linewidth}
        \includegraphics[width=\linewidth]{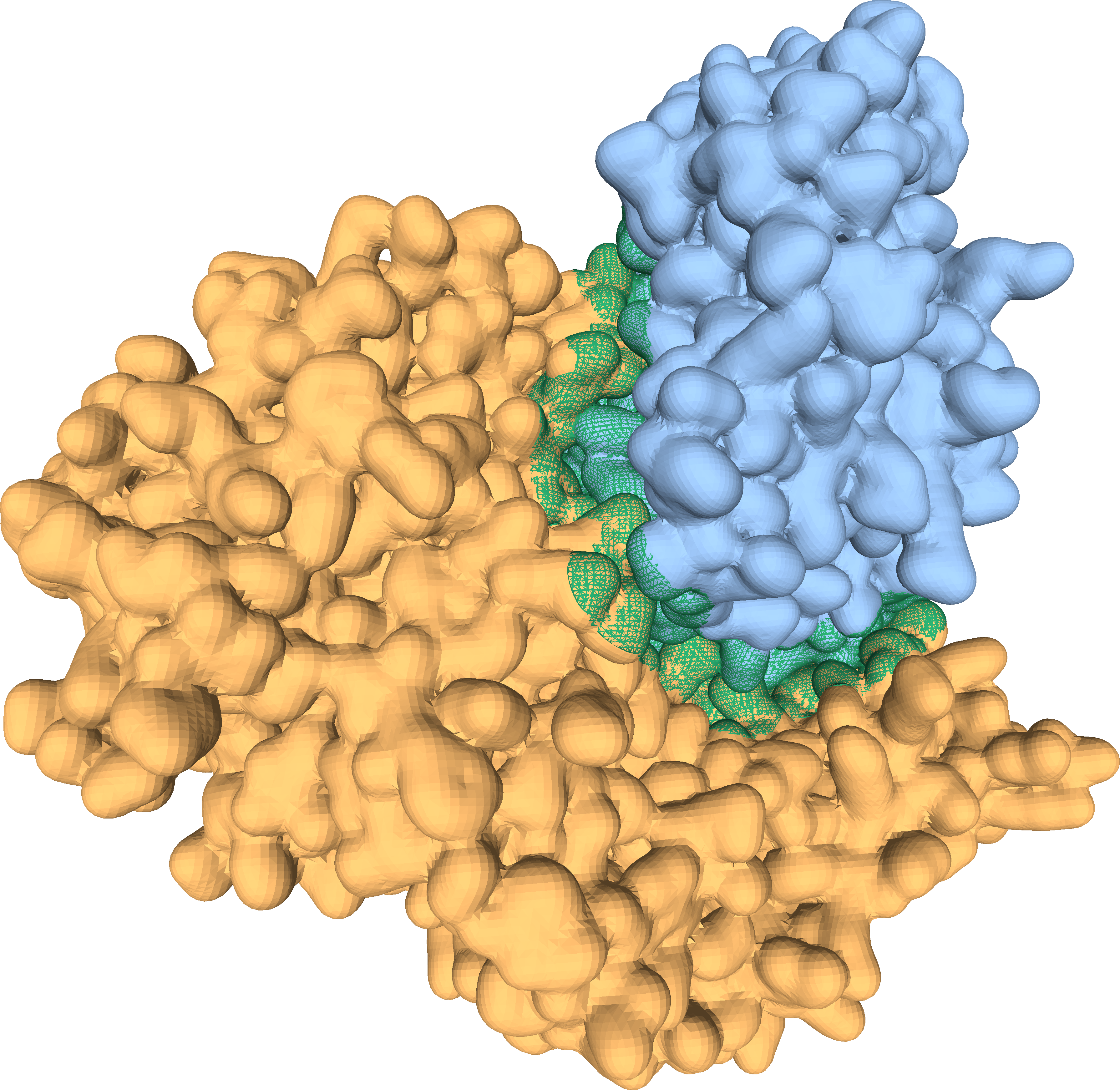}
        \caption{}
        \label{subfig:2gaf_both}
    \end{subfigure} 
    \hfill
    \begin{subfigure}[c]{.28\linewidth}
        \includegraphics[width=\linewidth]{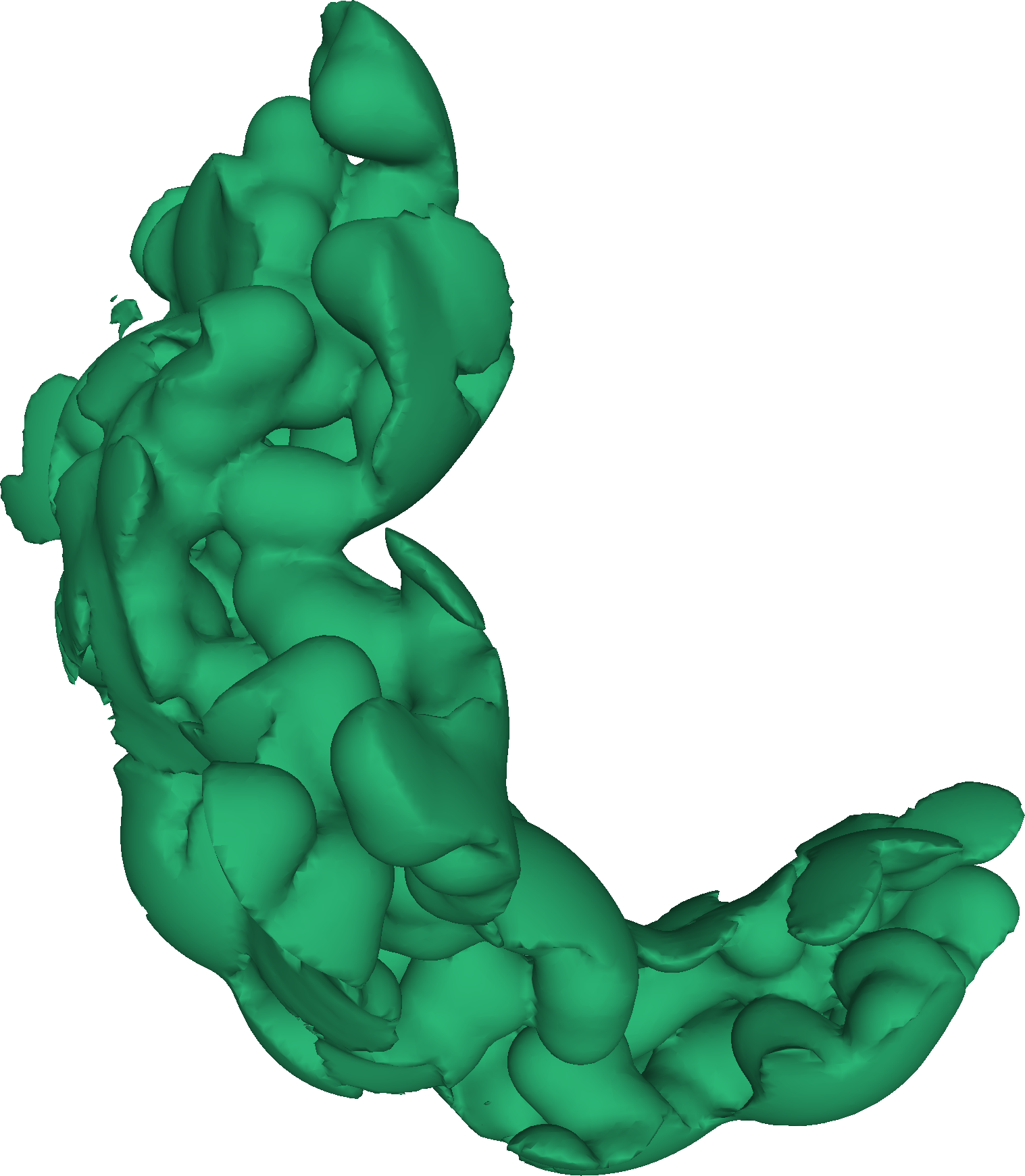}
        \caption{}
        \label{subfig:2gaf_inter_01}
    \end{subfigure}
    \hfill
    \begin{subfigure}[c]{.31\linewidth}
        \includegraphics[width=\linewidth]{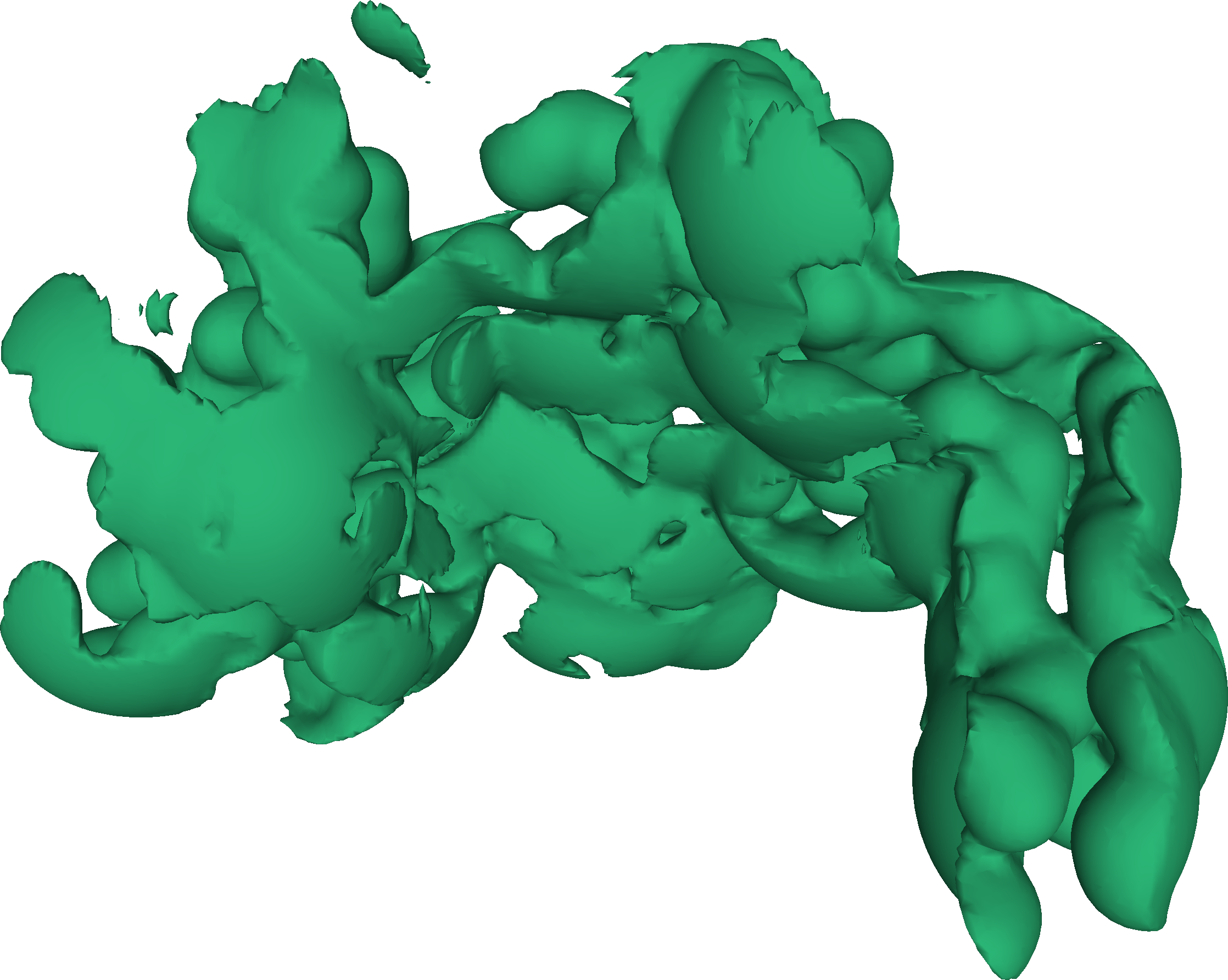}
        \caption{}
        \label{subfig:2gaf_inter_02}
    \end{subfigure}
    \hfill \null
    \caption{As in Figure \ref{fig:prot_intersect_ex1}, but for protein 2GAF. }
    \label{fig:prot_intersect_ex2}
\end{figure}

\begin{figure}
    \null \hfill
    \begin{tabular}{|l|l|l|}
    \hline
    & Avg time (ms) & Peak memory \\
    \hline \hline 
    Basic  & 48.01 & 1.91MB \\
    \hline
    KDTree & 77.43 & 11.15 kB \\
    \hline
    BVH & 45.77 & 7.06 kB \\
    \hline
  \end{tabular}
      \hfill \null
  \captionof{table}{Comparison of time/memory usage by two acceleration structures.}
  \label{tab:arch_compare}
\end{figure}

In this section, we use properties of SDFs to produce an implicit function that represents the shared interface between two protein chains. In addition to the smooth max defined in Equation \ref{eqn:smooth_min}, we make use of three operations on SDFs: 1) the rounding operation $d^{(r)}(\Vec{x}) = d(\Vec{x}) - r$, which expands the zero level set of an SDF by $r$ in the positive direction; 2) the intersection operation, where the intersection of two SDFs $d_1$ and $d_2$ is given by $d_{d_1 \cap d_2}(\Vec{x}) = \max \left( d_1(\Vec{x}), d_2(\Vec{x}) \right) $, and finally 3) the union operation, where the union of two SDFs $d_1$ and $d_2$ is given by $d_{d_1 \cup d_2}(\Vec{x}) = \min \left( d_1(\Vec{x}), d_2(\Vec{x}) \right) $. Let $d_1$ and $d_2$ be the SDFs for two protein chains. Then we calculate an interface SDF as:

\begin{align}
    d_{\textsc{INTER}} = \min\left( \right.
         &\max\left.\left( d_1\left(\Vec{x}\right) - r, d_2\left(\Vec{x}\right)  \right), \right. \label{eqn:intersection} \\
        &\max\left.\left( d_1\left(\Vec{x}\right), d_2\left(\Vec{x}\right) - r \right) \right) \nonumber
\end{align}
This produces an SDF of the parts of chain 1 that are within $r$ of chain 2, and vice versa. See Figures \ref{fig:prot_intersect_ex1} and \ref{fig:prot_intersect_ex2} for examples of these interface meshes with an intersection radius of $r = 4\text{\r{A}}$.

\subsection{Accelerating Queries}
\label{subsec:acceleration}
We investigate the use of two data structures for accelerating spatial queries of protein SDFs: Bounding Volume Hierarchies (BVHs) and K-D Trees. For further details about these data structures, we refer the reader to \cite{reinhard2000dynamic, foley2005kd, meister2021survey}. BVHs in particular have been previously explored as a technique to accelerate SDF queries \cite{liu2013exact}. Both of these data structures use spatial subdivision to accelerate finding the $k$ nearest points to a given query point. See Figure \ref{fig:bvh_with_bounding_boxes} for an illustration of the BVH approach. Both of these data structures yield an approximation (see Figure \ref{fig:acc_struct}) of the original signed distance function, but with a substantially lower memory footprint, which we analyze in Section \ref{subsec:arch_compare}.


\newcommand{\twodvizwidth}{0.43\linewidth}
\begin{figure}
    \centering
    \null \hfill \begin{subfigure}[b]{\twodvizwidth}
        \includegraphics[width=\textwidth]{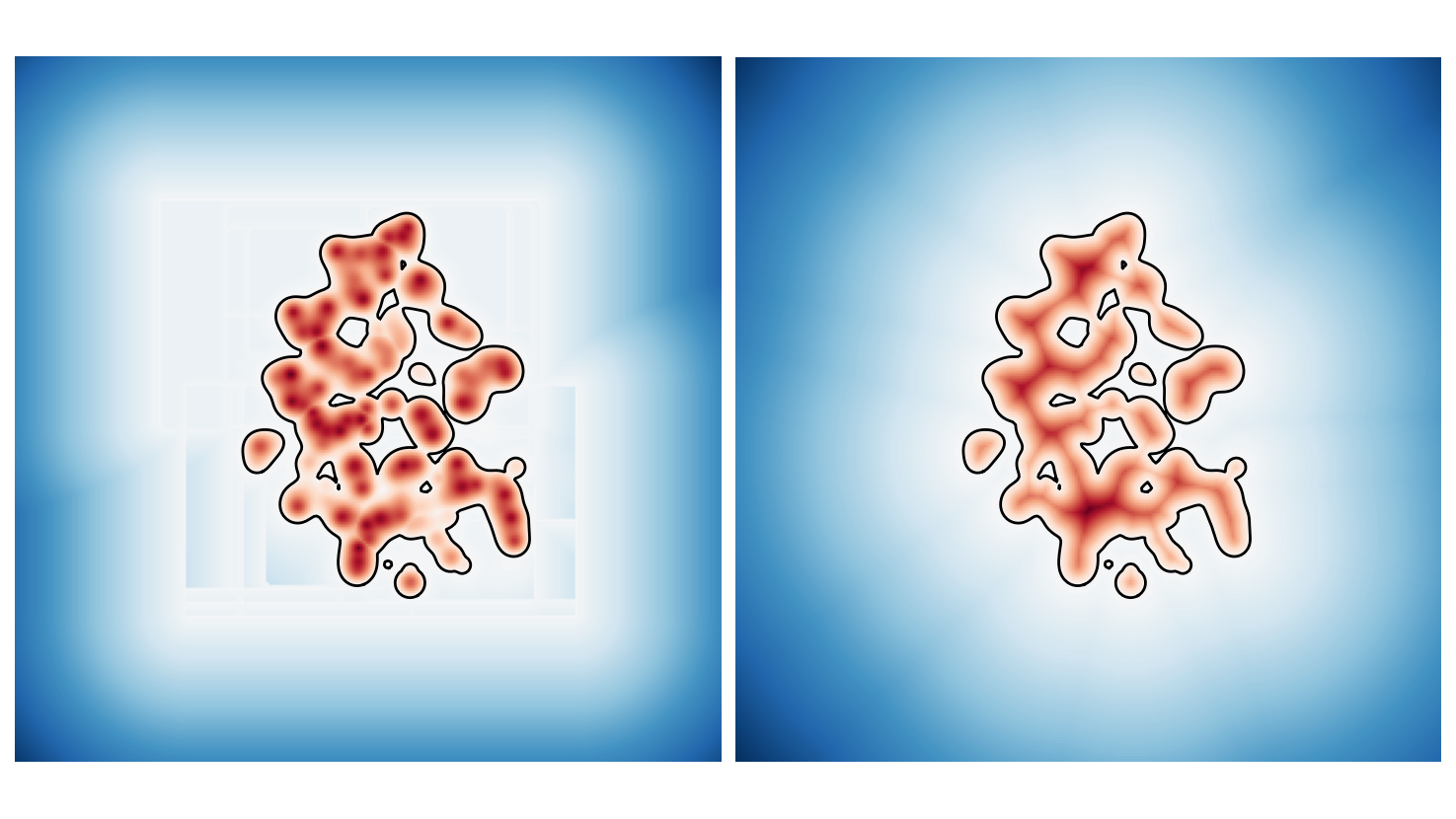}
        \caption{1A2K, BVH rendering. }
    \end{subfigure} \hfill
    \begin{subfigure}[b]{\twodvizwidth}
        \includegraphics[width=\textwidth]{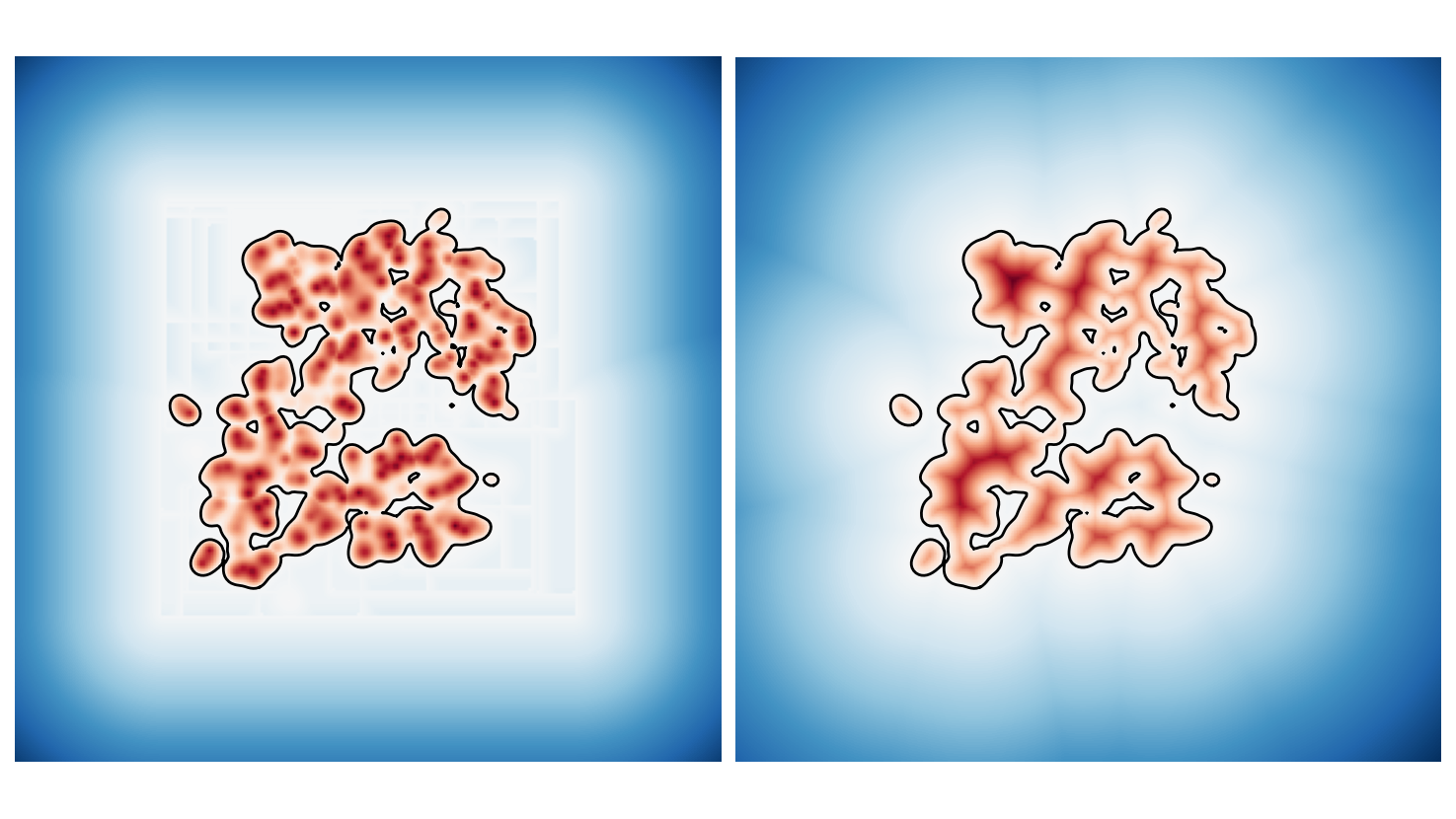}
        \caption{1WDW, BVH rendering. }
    \end{subfigure} \hfill \null \\
    \null \hfill \begin{subfigure}[b]{\twodvizwidth}
        \includegraphics[width=\textwidth]{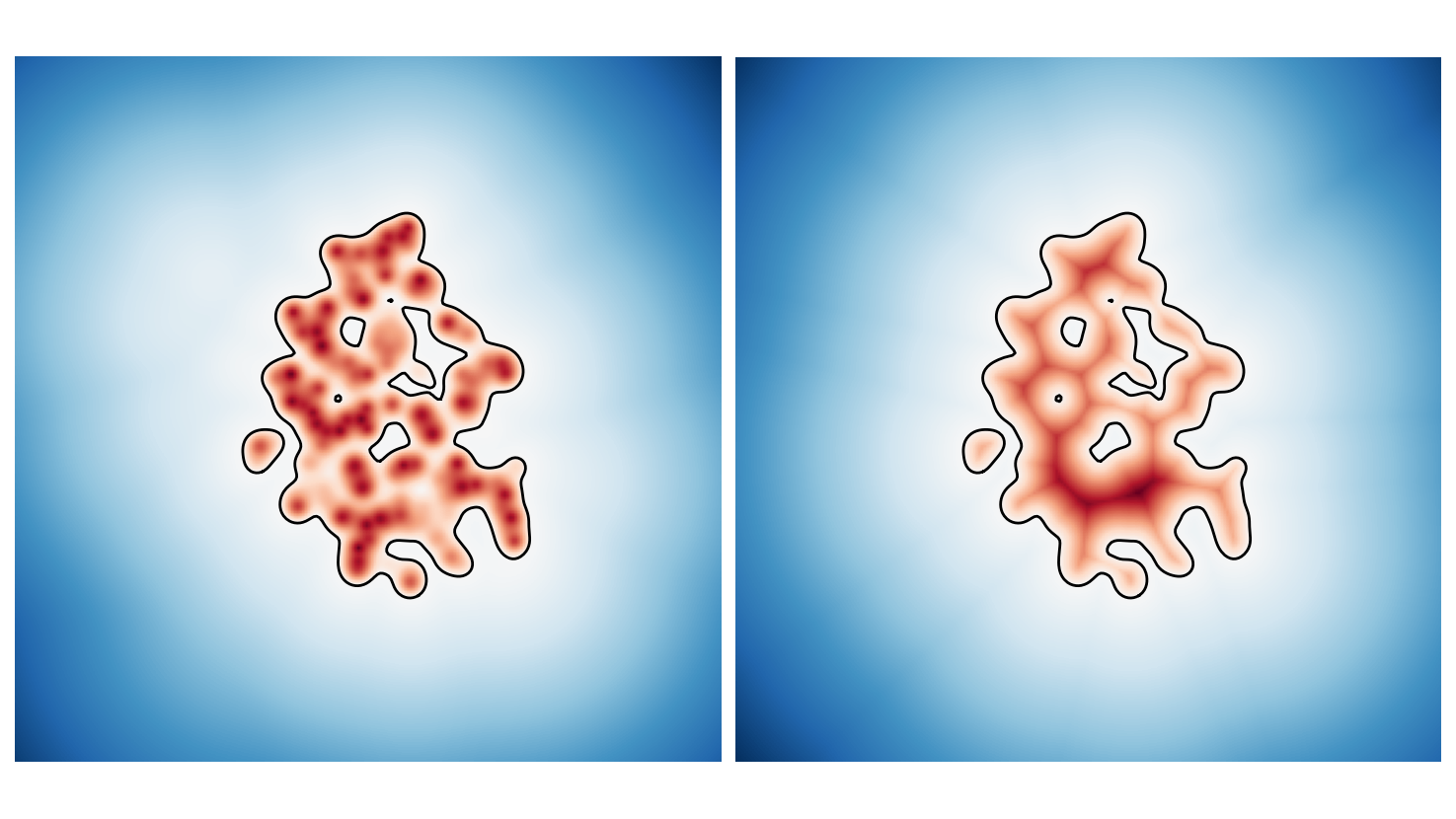}
        \caption{1A2K, kd-tree rendering. }
    \end{subfigure} \hfill
    \begin{subfigure}[b]{\twodvizwidth}
        \includegraphics[width=\textwidth]{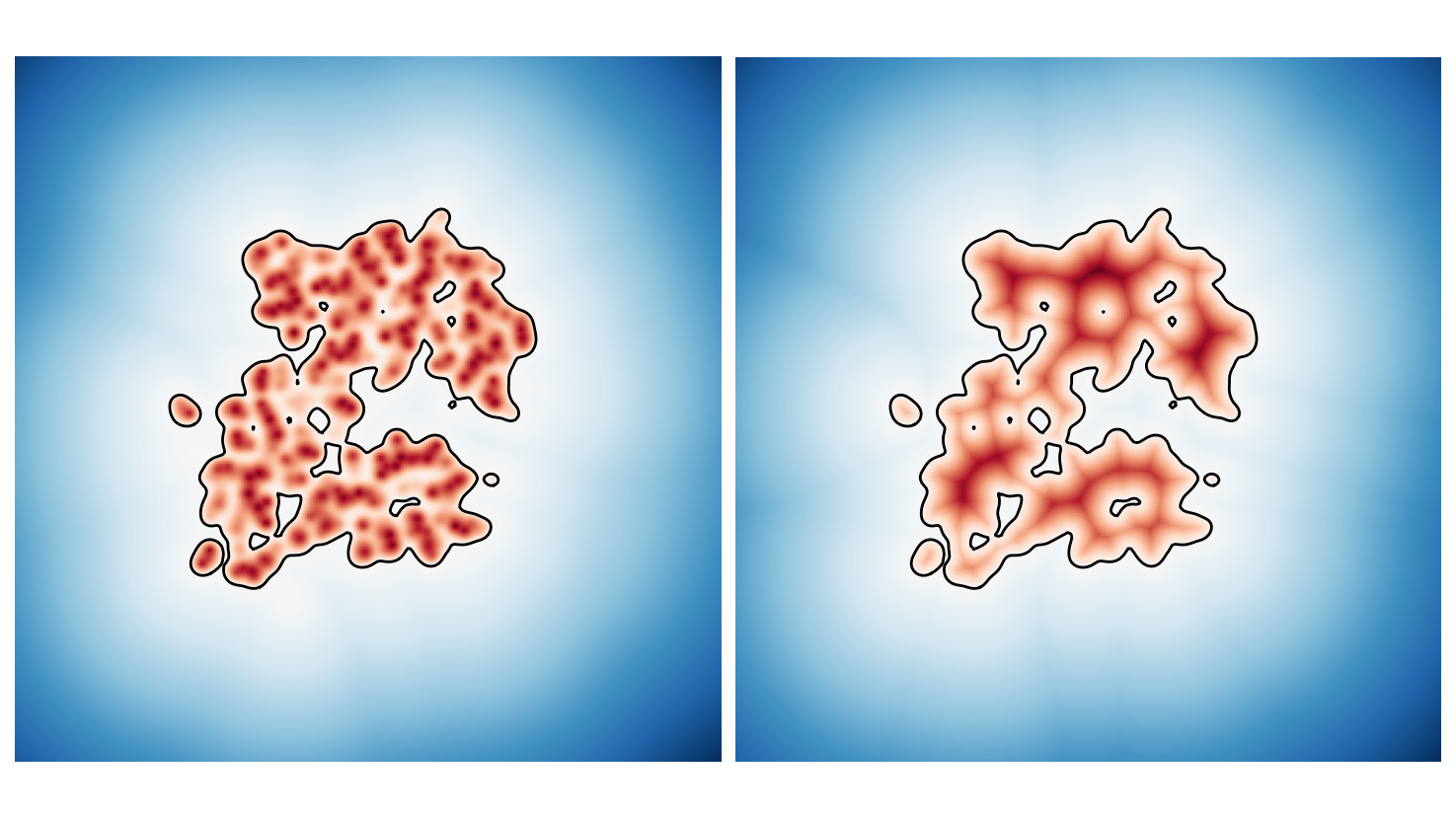}
        \caption{1WDW, kd-tree rendering. }
    \end{subfigure} \hfill \null
    \caption{Illustration of SDFs produced by the two acceleration structures we examine in this paper. Each picture shows a 2D slice through the signed distance field at $z=0$. We see that both accelerated SDFs are only approximations of the SDFs in Figure \ref{fig:flat_prot_figures}. In particular, the BVH approach leads to rectilinear artifacts on the exterior of the SDF; this could likely be ameliorated by using a different shape as the bounding volume (as opposed to axis-aligned rectangular prisms).}
    \label{fig:acc_struct}
\end{figure}

\subsection{Implementation Details}
Code used to build the SDFs described in this paper is available at \url{https://anonymous.4open.science/r/protein_sdfs-2E2C/README.md}. Our code is built on top of the \texttt{torch\_sdf} package, which is available at \url{https://anonymous.4open.science/r/torch_sdf-FFCB/README.md}. For visualizing SDFs, we first convert each to a mesh using the Marching Cubes algorithm \cite{lewiner2003efficient} as implemented in SciKit-Image \cite{van2014scikit}. We render images of each mesh using the \texttt{meshplot} package. 

\section{Results}
For this paper, we evaluate our model on the Docking Benchmark 5 (DB5) dataset \cite{db5paper}. This dataset consists of protein complexes, where each complex includes two chains. Each chain is represented with PDB files \cite{berman2000protein} with atom coordinates for both its undocked pose (i.e. the pose it folds into naturally) and its docked pose with the other protein in the complex. We use the ``bound" version of each chain in the complex. 
\subsection{Calculation of Protein Interaction Surfaces Using Smooth-Min SDFs}
We use the SDF defined in Equation \ref{eqn:intersection} to compute signed distance functions for all of the protein complexes in the DB5 dataset. We use an interaction radius of 4.0\r{A} in line with previous work on protein-protein interaction \cite{salamanca2017optimal,laskowski2018pdbsum}. See Figures \ref{fig:prot_intersect_ex1} and \ref{fig:prot_intersect_ex2} for example protein interface meshes. We provide interface meshes for all of the complexes in the DB5 dataset.
\subsection{Comparison of Acceleration Structures}
\label{subsec:arch_compare}
\begin{figure}
    \centering
    \includegraphics[width=0.47\linewidth]{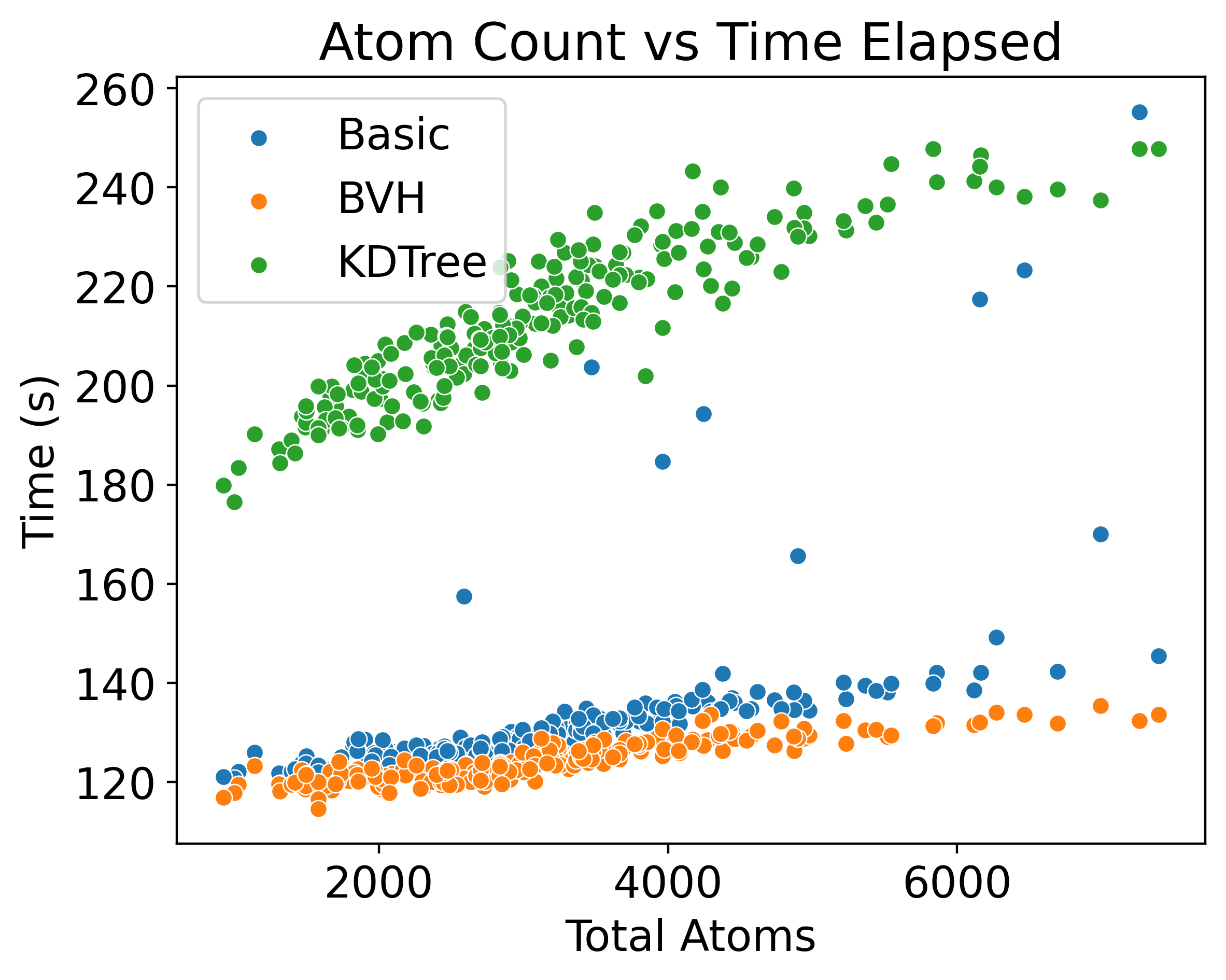} \hfill
    \includegraphics[width=0.47\linewidth]{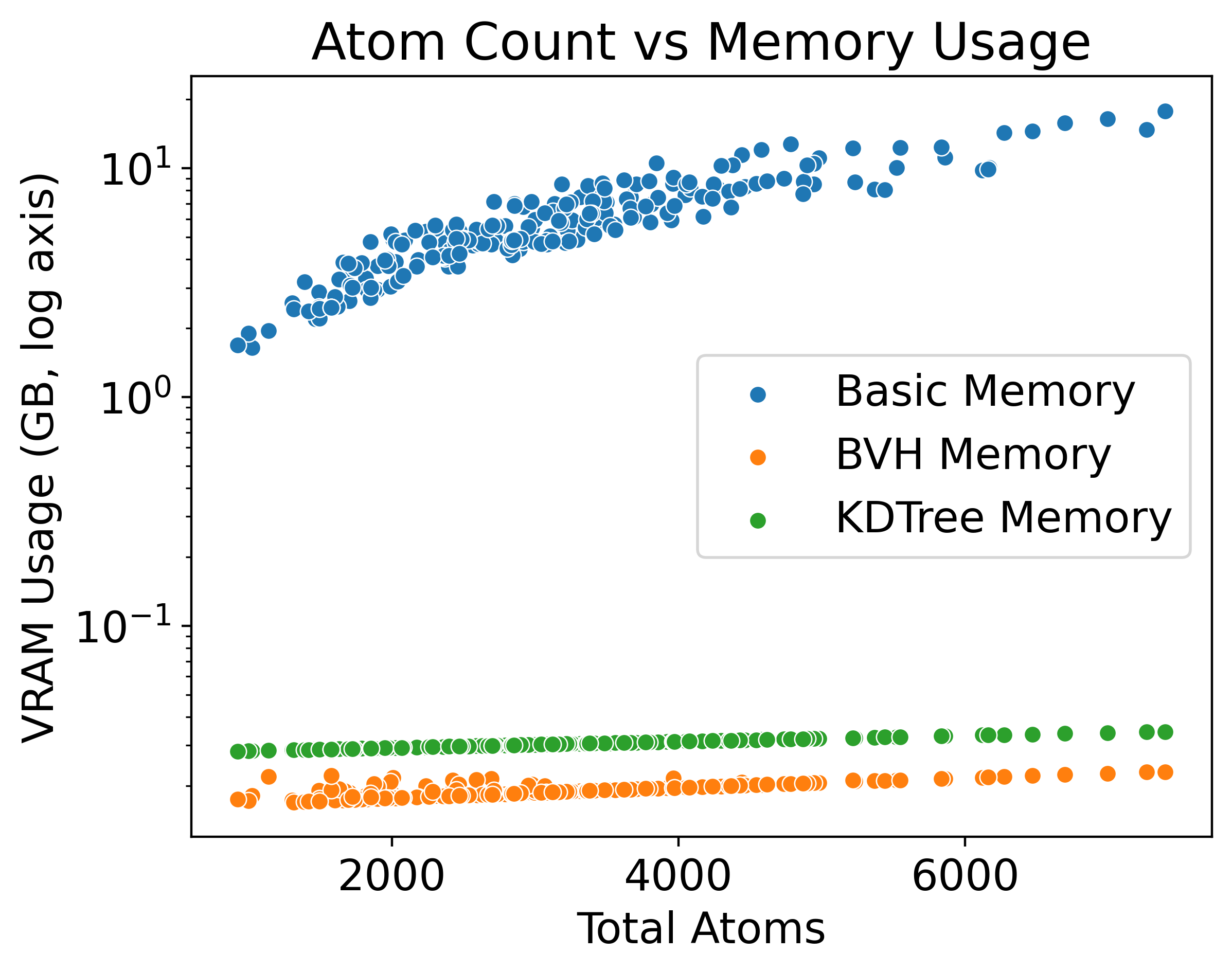} 
    \caption{Time/memory benchmarking of acceleration structures for querying protein SDFs. }
    \label{fig:acc_struct_compare}

\end{figure}

We compare the performance of the two acceleration structures described in Section \ref{subsec:acceleration}. For each protein complex in the DB5 dataset, we benchmark the time and GPU memory needed to evaluate all grid points in a $256^3$ grid laid over the bounding box of the complex. In Figure \ref{fig:acc_struct_compare}, we compare the accelerated version of each protein structure to the time and memory required to query all of the atomic SDFs for a given complex. We see that the BVH approach is slightly faster than the naive approach, but the KDTree is much slower, likely due to CPU-GPU latency (at time of writing, there was no functional Torch+CUDA KDTree implementation, so the KDTree approach necessitates copying data back to the CPU. We hope to ameliorate this inefficiency in future work). Both the KDTree and BVH approaches offer significant memory savings (see Table \ref{tab:arch_compare}). However, this approximation does incur some rectangular artifacts in the BVH tree approach; this could likely be resolved by using a different shape primitive in the volume hierarchy (e.g. bounding spheres instead of axis-aligned rectangles).

\section{Conclusion and Future Work}
This paper demonstrates the utility of representing a protein using the signed distance to its solvent-accessible surface (as approximated by the smooth-min function to the protein's component atoms). We also show how this can facilitate constructive solid geometry operations on protein interfaces. While this technique is promising, further work is needed to validate the proposed approach. One of the major advantages of building our protein representation in PyTorch \cite{paszke2017automatic} is the ability to optimize over parts of the protein representation. In future work, we hope to use our framework to optimize protein shape, conformation, and position according to SDF-based loss functions.

Recent work has used neural networks as maps between vectors of latent states and signed distance functions \cite{park2019deepsdf}, producing a single model that is able to predict SDF values for a dataset of multiple 3D shapes. We hope to investigate a combination of this approach and our SDF construction, perhaps by utilizing embedding vectors from a pretrained protein transformer architecture \cite{bio_embed}. We also hope to incorporate atomic characteristics as features of the generated surfaces. Finally, the literature includes examples of data structures like the ChainTree \cite{lotan2003efficient} which have been specifically developed for fast querying of energy potentials of proteins. It is likely that our proposed approach could be further enhanced by one of these protein-specific acceleration structures. 

Finally, we note that constructing an SDF that implicitly stores the SAS shape is only the first step. We anticipate that this approach could be useful in any case where a machine learning model operates on the solvent accessible surface (AKA, where differentiability is needed). An example could include optimizing the position and pose of amino acids to produce an SAS that fits some other structural motif; or to provide some other machine learning model with a training signal that takes protein shape into account. We hope to use our SAS representation in one or more of these applications. 

\newpage

\bibliographystyle{ws-procs11x85} 
\bibliography{references}

\end{document}